\documentclass{article}
\usepackage{graphicx}

\begin{document}

\large

\title{\bf Perturbation Theory for the Double--Sine--Gordon Equation}%
\author{Constantine A. Popov\\
\small Volgograd State Pedagogical University, Lenina Prospect 27, Volgograd 400131, Russia\\
\small e-mail: popovca@yandex.ru}

\maketitle
% ----------------------------------------------------------------
\begin{abstract}
This paper presents the perturbation theory for the
double--sine--Gordon equation. We received the system of
differential equations that shows the soliton parameters
modification under perturbation's influence. In particular case
$\lambda=0$ the results of the research transform into well-known
perturbation theory for the sine--Gordon equation.
\end{abstract}


\noindent \textbf{Pacs Numbers}: 02.30.Jr, 05.45.Lj.

% ----------------------------------------------------------------
\section*{Introduction}

\hspace*{\parindent}The double--sine--Gordon (DSG) equation  was
studied in some physical problems. It describes spin waves in
superfluid 3He, self-induced transparency in accounting degeneracy
of atomic levels \cite{salamo}, electromagnetic waves propagation
in semiconductor quantum supelattices \cite{kruchkov}, etc. The
unperturbed DSG-equation can be written as

\begin{equation}\label{udsg}
    {{\partial^2 \varphi} \over {\partial t^2}} - {{\partial^2 \varphi} \over {\partial
    x^2}} + \sin \varphi + \lambda \sin 2 \varphi = 0.
\end{equation}

DSG-equation can not be solved with the inverse scattering
transform. But it is known that equation (\ref{udsg}) has the kink
solution

\begin{equation}\label{sol}
    \varphi = 2 \arctan \left( \sqrt{2\lambda + 1} \sinh^{-1} \left( {\sqrt{2\lambda + 1} \over
    \sqrt{1 - u^2} }(x-x_0-ut) \right) \right).
\end{equation}

We shall call this solution as DSG-soliton, to differ it from the
soliton of the sine-Gordon (SG) equation (particular case of the
DSG-soliton, when $\lambda=0$). Interactions of solitary waves
like (\ref{sol}) are not elastic \cite{ablowitz, bullough, olsen},
they are accompanied by the radiation loss. The propagation media
also influences the soliton parameters (its speed $u$ and its
location $x_0$). This influence can be described as perturbation
in (\ref{udsg}). Thus, we have

\begin{equation}\label{pdsg}
    {{\partial^2 \varphi} \over {\partial t^2}} - {{\partial^2 \varphi} \over {\partial
    x^2}} + \sin \varphi + \lambda \sin 2 \varphi = \varepsilon f.
\end{equation}

The solution (\ref{sol}) does not exactly satisfy (\ref{pdsg}).
But if the perturbation is small we can take the DSG-soliton with
slowly varying parameters as the solution. Then we find equations,
which describe modifications of $u$ and $x_0$. For this purpose we
construct the perturbation theory.

\section*{Perturbation theory (main algorithm)}

\hspace*{\parindent}This paper is not the first attempt to solve
the perturbed DSG-equation. This problem was examined in
assumption $\lambda \ll 1$ \cite{mason} using the inverse
scattering transform for the perturbed sine--Gordon equation. But
the paper did not show the DSG-soliton parameters modification.

There is a number of methods for the perturbation theory
constructing \cite{yan}-\cite{maniadis}. As it was shown in
\cite{salerno} the perturbation theory for the SG-equation by
McLaughlin and Scott describes the experimental results and the
numerical solutions of the SG-equation better. Therefore, we use
the algorithm represented in \cite{lonngren}. For these purposes
we modify and adopt this algorithm for the DSG-equation.

Let's rewrite the DSG-equation in matrix form

\begin{equation}\label{matr_dsg}
    \partial_t \left[ \begin{array}{c} \varphi \\ \varphi_t \\ \end{array}
    \right] + \left[ \begin{array}{cc} 0 & -1 \\ -\partial_{xx} + \sin (\circ) + \lambda \sin (2 \ \circ) &
    0 \\ \end{array} \right] \left[ \begin{array}{c} \varphi \\ \varphi_t \\ \end{array}
    \right] = \varepsilon \left[ \begin{array}{c} 0 \\ f(\varphi) \\
    \end{array} \right].
\end{equation}

We mark the column $(\varphi, \varphi_t)$ as $W$

\begin{equation}
    W = \left[ \begin{array}{c} \varphi \\ \varphi_t \\
    \end{array} \right].
\end{equation}

$W$ is the function of the DSG-soliton parameters $u$ and $x_0$.
So, $W = W(u, x_0)$ or $W = W(p)$, where $p = (u, x_0)$. We
suppose that the solution of the equation (\ref{pdsg}) can be
presented as

\begin{equation}\label{pert}
    W = W_0 + \varepsilon W_1.
\end{equation}

\noindent Here $W_0$ is the exact solution of (\ref{udsg}) with
the dependent on time parameters.

If we insert this perturbed solution into (\ref{matr_dsg}) and
linearize it, then we rewrite it in the following form

\begin{equation}\label{oper_form}
    L(W_1) = F(W_0).
\end{equation}

\noindent Here $L$ is the linear operator which is given as

\begin{equation}\label{l}
    L = \left[ \begin{array}{cc} 1 & 0 \\ 0 & 1 \\ \end{array} \right] \partial_t +
    \left[ \begin{array}{cc} 0 & -1 \\ -\partial_{xx} + \cos \varphi_0 + 2 \lambda \cos 2 \varphi_0 &
    0 \\ \end{array} \right].
\end{equation}

The right part of (\ref{oper_form}) represents the 'effective
force'

\begin{equation}\label{force}
    F(W_0) = f(W_0) - {1 \over \varepsilon} \sum_{i=1}^2 {{\partial
    p_i} \over {\partial t}} {{\partial W_0} \over {\partial
    p_i}}.
\end{equation}

The boundedness condition for $W_1$ is provided with the
orthogonality $F$ to the discrete subspace of the kernel
$N_d(L^+)$, where $L^+$ is the adjoint operator to $L$

\begin{equation}
    L^+ = -\left[ \begin{array}{cc} 1 & 0 \\ 0 & 1 \\ \end{array} \right] \partial_t +
    \left[ \begin{array}{cc} 0 & -\partial_{xx} + \cos \varphi_0 + 2 \lambda \cos 2 \varphi_0 \\ -1 &
    0 \\ \end{array} \right].
\end{equation}

Inserting the basis in $N_d(L^+)$ and putting it into
(\ref{force}) we derive the equation, that satisfies the
orthogonality condition

\begin{equation}\label{base0}
    \sum_{j=1}^2 \left( b_i, {{\partial W_0} \over {\partial p_j}}
    \right){{d p_j} \over {d t}}= \varepsilon (b_i, f), \ \ \ i=1,
    2,
\end{equation}

\noindent where $ \left( F(x), G(x) \right) = \int_{-
\infty}^{\infty} F^T(x) G(x) dx$. And $b_i$ are the basis vectors
of the adjoint space

\begin{equation}
    b_i = \left[ \begin{array}{cc} 0 & -1 \\ 1 & 0 \\ \end{array}
    \right] {{d W_0} \over {d p_i}}.
\end{equation}

The equation (\ref{base0}) represents the perturbation theory for
the DSG-soliton parameters.

\section*{Analysis of the DSG-soliton motion}

\hspace*{\parindent}And now let's consider the case of the
DSG-soliton propagation. We have

\begin{equation}
    {{dW_0} \over {dx}} = \left[ \begin{array}{c} \varphi_x \\ \varphi_{tx} \\
    \end{array} \right], \ \ \ {{dW_0} \over {du}} = \left[ \begin{array}{c} \varphi_u \\ \varphi_{tu} \\
    \end{array} \right].
\end{equation}

Thus, we receive the basis vectors of the adjoint space as

\begin{equation}
    b_1 = \left[ \begin{array}{c} -\varphi_{tx} \\ \varphi_x \\
    \end{array} \right], \ \ \ b_2 = \left[ \begin{array}{c} -\varphi_{tu} \\ \varphi_u \\
    \end{array} \right].
\end{equation}

It allows us rewriting the system (\ref{base0})

\begin{equation}\label{base1}
    \left\{ \begin{array}{l}
      \left(
      \left[\begin{array}{c} -\varphi_{tx} \\ \varphi_x \\ \end{array} \right],
      \left[\begin{array}{c} -\varphi_u \\ \varphi_{tu} \\ \end{array} \right]
      \right)
      \displaystyle\frac{du}{dt} = \varepsilon
      \left(
      \left[\begin{array}{c} -\varphi_{tx} \\ \varphi_x \\ \end{array} \right],
      \left[\begin{array}{c} 0 \\ f(\varphi_0) \\          \end{array} \right]
      \right), \\
       \\
      \left(
      \left[\begin{array}{c} -\varphi_{tu} \\ \varphi_u \\ \end{array} \right],
      \left[\begin{array}{c} -\varphi_x \\ \varphi_{tx} \\ \end{array} \right]
      \right)
      \displaystyle\frac{dx_0}{dt} = \varepsilon
      \left(
      \left[\begin{array}{c} -\varphi_{tu} \\ \varphi_u \\ \end{array} \right],
      \left[\begin{array}{c} 0 \\ f(\varphi_0) \\          \end{array} \right]
      \right). \\
    \end{array} \right.
\end{equation}

The other summands in (\ref{base0}) equal to zero. If we transfer
these equations into integral form we receive

\begin{equation}\label{base2}
    \left\{ \begin{array}{l}
      \left( \displaystyle\int_{-\infty}^{\infty} (\varphi_{tu} \varphi_x -
      \varphi_{tx} \varphi_u ) dx \right)
      \displaystyle\frac{du}{dt} = \varepsilon
      \displaystyle\int_{-\infty}^{\infty} f(\varphi_0) \varphi_x dx , \\
       \\
      \left( \displaystyle\int_{-\infty}^{\infty} (\varphi_{tx} \varphi_u -
      \varphi_{tu} \varphi_x ) dx \right)
      \displaystyle\frac{dx_0}{dt} = \varepsilon
      \displaystyle\int_{-\infty}^{\infty} f(\varphi_0) \varphi_u dx. \\
    \end{array} \right.
\end{equation}

These equations transform into a system of two equations. They
describe the evolution of soliton parameters, its speed and
location.

\begin{equation}\label{system}
    \left\{ \begin{array}{l}
    \displaystyle\frac{du}{dt} = -\varepsilon \displaystyle\frac{1-u^2}{2} \frac{2 \lambda
    +1}{k(\lambda)} \int_{-\infty}^{\infty} f(\varphi_0)
    \displaystyle\frac{\cosh(\theta)}{\cosh^2(\theta) + 2 \lambda} dx, \\
      \\
    \displaystyle\frac{dx_0}{dt} = u - \varepsilon \displaystyle\frac{u \sqrt{1-u^2}}{2} \frac{2 \lambda
    +1}{k(\lambda)} \int_{-\infty}^{\infty} f(\varphi_0)
    \displaystyle\frac{\theta \cosh(\theta)}{\cosh^2(\theta) + 2 \lambda} dx. \\
    \end{array} \right.
\end{equation}

\noindent Here $\theta = \sqrt{\displaystyle\frac{2 \lambda +
1}{1-u^2}} (x-x_0-ut)$, $k(\lambda) = \sqrt{2 \lambda +1} +
\displaystyle\frac{1}{\sqrt{2 \lambda}} \tanh^{-1}\left(
\sqrt{\displaystyle\frac{2 \lambda}{2 \lambda + 1}} \right)$.

If we put $\lambda = 0$, then $k(0)=2$ and this system reduces to
a similar system for the perturbed SG-equation \cite{lonngren}.
It's easy to notice, that the parameters modification depends on
the value of $\lambda$ and the type of perturbation function
$f(\varphi_0)$. This function can be written in different
variations. But the most frequently appeared perturbation function
contains three summands

\begin{equation}\label{pert_func}
    f(\varphi) = - \gamma - \alpha \frac{\partial \varphi}{\partial
    t}- \mu [\sin(\varphi) + \lambda \sin(2 \varphi)] \delta (x).
\end{equation}

The influence of the constant energy pumping (the summand of
$\gamma$), the energy losses (the summand with $\alpha$) and the
presence of the inhomogeneous region (like the microshort in the
Josephson junctions) are represented in this perturbation
function. If we take $\mu = 0$ and substitute the expression of
(\ref{pert_func}) in (\ref{system}), we shall get the following
system of differential equations

\begin{equation}\label{sys1}
    \left\{ \begin{array}{l}
    \displaystyle\frac{du}{dt} = \displaystyle\frac{\pi \gamma}{2 k(\lambda)} \left( 1 - u^2
    \right)^{3/2} - \alpha u \left( 1 - u^2 \right), \\
    \\
    \displaystyle\frac{dx_0}{dt} = u.\\
    \end{array} \right.
\end{equation}

Analyzing this system we derive the speed stability condition. To
satisfy the condition the right part of the first equation must be
equal to zero

\begin{equation}\label{equilibr}
\displaystyle\frac{\pi \gamma}{2 k(\lambda)} \left( 1 - u^2
    \right)^{3/2} - \alpha u \left( 1 - u^2 \right) = 0.
\end{equation}

So, we get the stabilized speed value

\begin{equation}\label{u_stab}
    u_0 = \frac{1}{\sqrt{1 + \left( \displaystyle\frac{2 k(\lambda) \alpha}{\pi \gamma}
    \right)^2}}.
\end{equation}

This expression shows that if the energy pumping is minimal
($\gamma \rightarrow 0$) the soliton speed reduces to 0. On the
contrary if the DSG-soliton takes the energy much faster than it
looses ($\gamma \gg \alpha$) its speed increases up to 1. Putting
$\lambda = 0$ we take the expression for the stabilized speed of
the SG-soliton \cite{lonngren}. Any DSG-soliton reaches the speed
of $u_0$ under the action of energy pumping and energy loss
without the difference of its initial speed.

In case $\mu \neq 0$ the system (\ref{system}) can be written as

\begin{equation}\label{sys2}
    \left\{ \begin{array}{l}
      \displaystyle\frac{du}{dt} = \displaystyle\frac{\pi \gamma}{2 k(\lambda)} \left( 1 - u^2
      \right)^{3/2} - \alpha u \left( 1 - u^2 \right) + \mu \left( 1 - u^2 \right) g(\theta_0) \\
      \\
      \displaystyle\frac{dx_0}{dt} = u - \mu u x_0 \sqrt{2 \lambda + 1} g(\theta_0).\\
    \end{array} \right.
\end{equation}

\noindent Here $\theta_0 =
\sqrt{\displaystyle\frac{2\lambda+1}{1-u^2}}x_0$ and $g(x) =
\displaystyle\frac{(2\lambda+1)^{5/2}}{2 k(\lambda)}
\displaystyle\frac{\cosh^2(x)-2\lambda}{\left( \cosh^2(x)+2\lambda
\right)^3} \sinh(2x)$.

\begin{figure}[thb]
\begin{tabular}{cc}
    \includegraphics[width=75mm]{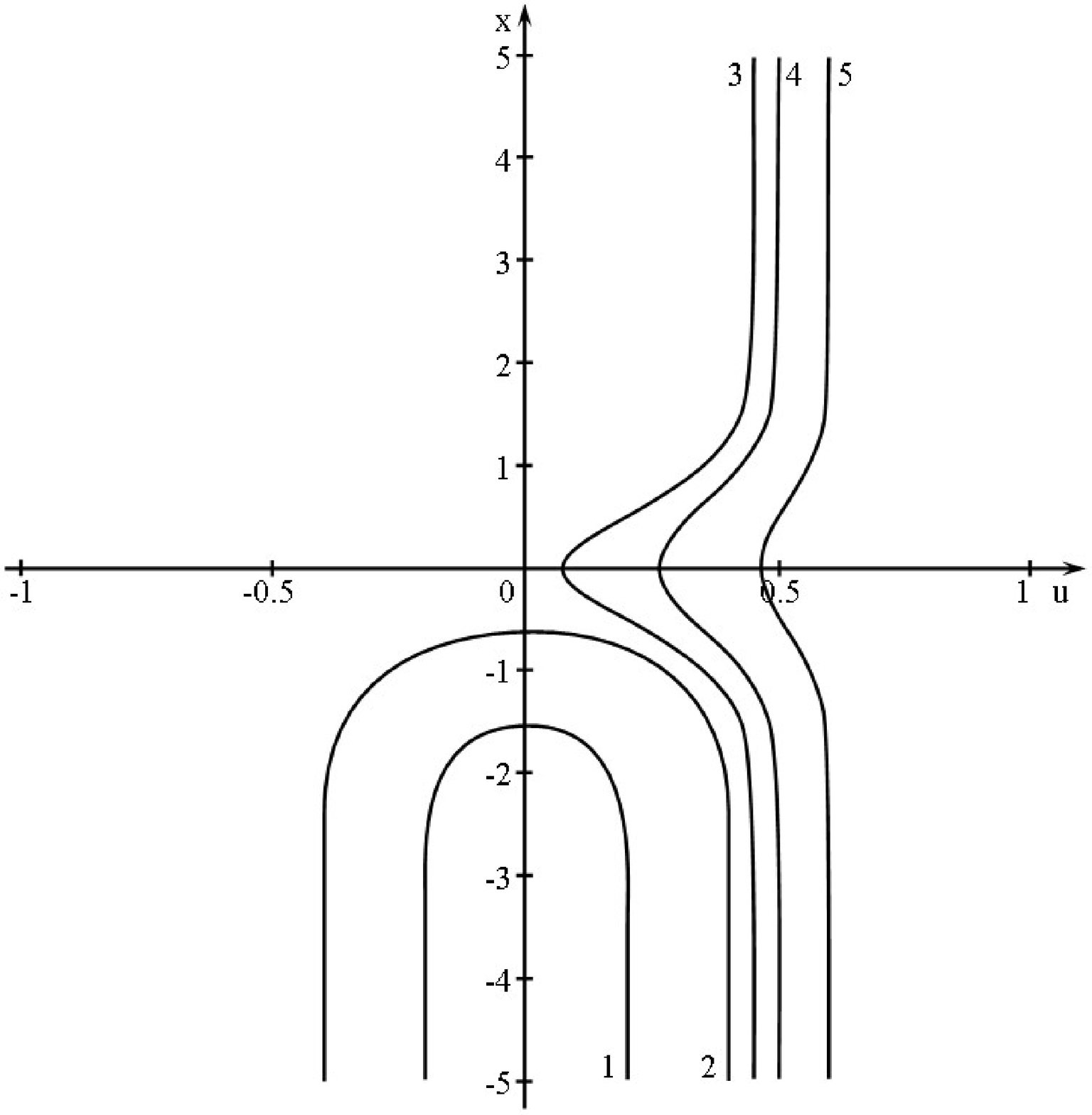} &
    \includegraphics[width=75mm]{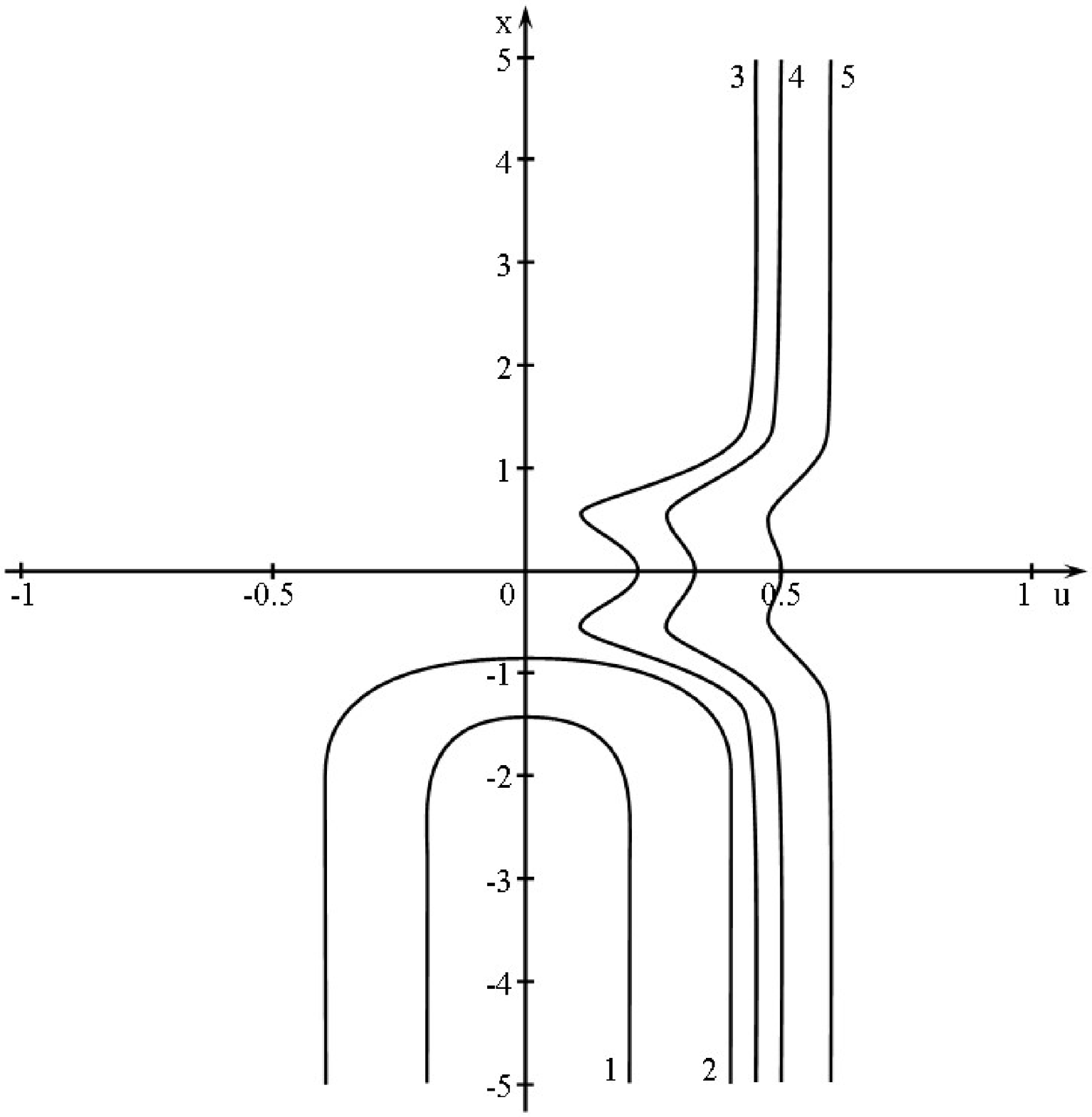}
\end{tabular}
\caption{The DSG-soliton interaction with the inhomogeneity. The
parameters are $\lambda = $ 0.2 (left), 1.2 (right); $\mu = 0.5$,
$\gamma = \alpha = 0$. The initial soliton speed:
  1) 0.2; 2) 0.4; 3) 0.45; 4) 0.5; 5) 0.6.}
\label{fig1}
\end{figure}

If we put the values $\gamma = 0$ and $\alpha = 0$, then we shall
get the case of the pure interaction between the DSG-soliton and
the inhomogeneity. The DSG-solitons with low speed reflects from
the inhomogeneity. And high speed solitons overcomes this region.
The results of interactions are shown on Figure \ref{fig1}.

It is possible to find the condition of overcoming from the energy
analysis. The Hamiltonian function of the unperturbed DSG-equation
is defined as

\begin{equation}\label{hamilton}
    H = \int_{-\infty}^{\infty} \left( \frac{\varphi_t^2}{2} +
    \frac{\varphi_x^2}{2} + (1 - \cos\varphi) + \frac{\lambda}{2} (1 -
    \cos2\varphi) \right) dx.
\end{equation}

Substituting (\ref{sol}) into (\ref{hamilton}) we receive the
expression for the full soliton energy

\begin{equation}\label{energy}
    E = \frac{4 k(\lambda)}{\sqrt{1 - u^2}}.
\end{equation}

To find the 'energy of rest' let's put the speed value which is
equal to 0. Thus, we get the kinetic energy of the DSG-soliton

\begin{equation}\label{kin_en}
    E_k = 4 k(\lambda) \left( \frac{1}{\sqrt{1 - u^2}}
    - 1 \right).
\end{equation}

The inhomogeneous area can accumulate energy. This process can be
counted as the correction to the DSG-Hamiltonian

\begin{equation}\label{ham_micro}
    H_i = \int_{-\infty}^{\infty} \mu \left( (1 - \cos\varphi) + \frac{\lambda}{2}
    (1 - \cos2\varphi) \right) \delta(x) dx.
\end{equation}

Having substituted (2) into the previous expression we get this
correction

\begin{equation}\label{micro0}
    H_i = 2 \mu (2 \lambda + 1)^2 \left( \frac{\cosh \theta_0}{\cosh^2 \theta_0 + 2 \lambda}
    \right)^2.
\end{equation}

The maximal energy which can be accumulated by the inhomogeneity

\begin{equation}\label{micro1}
    H_{max} = \left\{ \begin{array}{ll}
      2 \mu, & \mbox{if} \ -0.5 < \lambda < 0.5; \\
       &  \\
      \displaystyle\frac{(2 \lambda + 1)^2}{4 \lambda} \mu, & \mbox{if} \ \lambda > 0.5. \\
    \end{array}    \right.
\end{equation}

The value of the critical speed can be found from the condition of
equality of the kinetic energy (\ref{kin_en}) and the energy
accumulated in the inhomogeneity (\ref{micro1}).

\begin{equation}\label{u_crit}
    u_c = \sqrt{\frac{H_{max}^2 + 8 H_{max} k(\lambda)}{[H_{max} + 4
    k(\lambda)]^2}}.
\end{equation}

\begin{figure}[thb]
  \center \includegraphics[width=75mm]{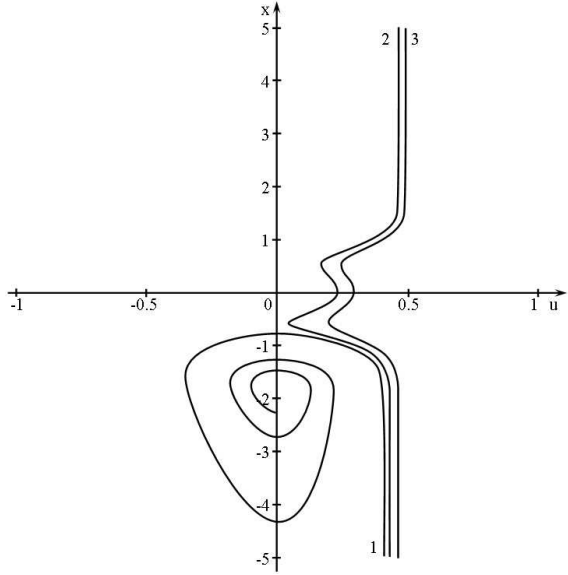}\\
  \caption{The DSG-soliton interaction with inhomogeneity in case $\alpha = 0.04$,
  $\mu = 0.5$, $\lambda = 1.2$, $\gamma$: 1) 0.030; 2) 0.032; 3) 0.035.}\label{fig2}
\end{figure}

The one more thing must be mentioned about the DSG-soliton
interaction with inhomogeneity. If all media parameters ($\gamma$,
$\alpha$, $\mu$) are not equal to zero the phase path is not
symmetric concerning the $u$-axis. This effect is shown on Figure
\ref{fig2}.

This asymmetry is caused by the energy pumping presence. As soon
as the soliton begins to loose its speed the equilibrium condition
(\ref{equilibr}) breaks. From this moment the system has the
positive balance of the energy gain. This process changes the
DSG-soliton speed and, therefore, the phase path changes its
shape. This effect takes place in the same problem for the
SG-equation, but usually it isn't considered. The asymmetry is
small if the media parameters $\alpha$, $\gamma \ll 1$ or the
soliton speed $u \rightarrow 1$.

In addition to this effect we notice the soliton damping after
reflection. The damping is caused by the influence of energy
inflow that pulls the solitary wave to the positive direction
along the $x$-axis.

\begin{figure}[thb]
  \center \includegraphics[width=75mm]{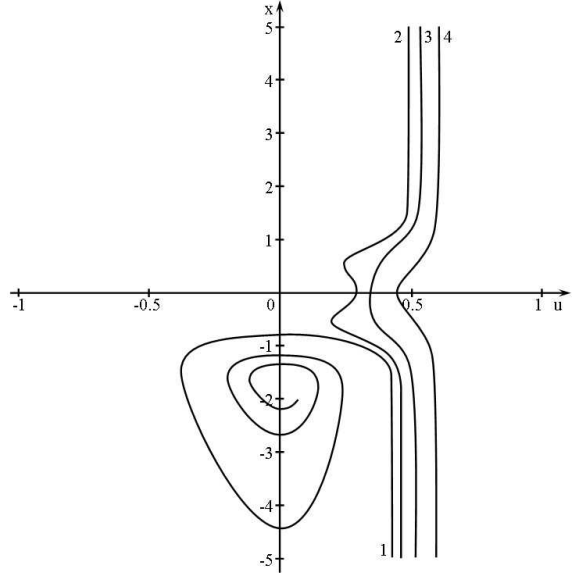}\\
  \caption{The influence of $\lambda$ on the DSG-soliton motion in case
  $\gamma = 0.035$, $\alpha = 0.04$, $\mu = 0.5$,
  $\lambda$: 1) -0.2; 2) 0.5; 3) 1.2; 4) 1.9.}\label{fig3}
\end{figure}

The numerical analysis shows that the view of the phase path
depends on the value of the $\lambda$-parameter (see Figure
\ref{fig3}).

It is necessary to notice that the condition (\ref{micro1}) allows
to classify the DSG-solitons. The DSG-soliton (exactly, its
derivative) looks like one-humped solitary wave (like the
SG-soliton), when $\lambda \in (-0.5, 0.5)$. If $\lambda > 0.5$,
the DSG-soliton is a two-humped wave. That's why the phase path
has two turning points if . And the phase path has the only one
extreme point in case $\lambda < 0.5$. Thus, it is possible to
evaluate the $\lambda$-parameter observing the soliton behaviour.
And vice versa. We can control the propagation process by changing
the value of $\lambda$.

\section*{Conclusion}

\hspace*{\parindent}In the conclusion we shall briefly observe
some results. First of all, this paper contains the perturbation
theory that has been adopted for the DSG-equation. The derived
formulas were used to consider the DSG-soliton propagation. It is
shown that the DSG-soliton can be stabilized in the presence of
the constant energy pumping and energy loss. We derive the
condition of the stabilization. Also the DSG-soliton interaction
with inhomogeneity was studied. We notice two variants of
interaction: reflection and overcoming. And so we receive the
condition of their separation. The variant of the propagation
depends on the parameters of the media: $\alpha$, $\gamma$, $\mu$
and $\lambda$.

In case $\lambda = 0$ all the results transform into the same
expressions for the SG-equation \cite{lonngren}.

\section*{Appendix A}

\hspace*{\parindent}We represent the integrals that were used in
the calculations. We use $\tanh^{-1}$ to mark the inverse
hyperbolic tangent. To derive these expressions well-known
integrals from \cite{dwight} were taken.

$$\displaystyle\int \displaystyle\frac{dx}{\cosh^2x + a} =
\displaystyle\frac{1}{\sqrt{a(a+1)}}\tanh^{-1}\left(
\sqrt{\displaystyle\frac{a}{a+1}}\tanh x \right).$$

$$ \begin{array}{c}
\displaystyle\int \displaystyle\frac{\cosh^2x \ dx}{(\cosh^2x +
a)^2} = - \displaystyle\frac{(2a+1) e^{2x}+1}{(a+1)
\left(e^{4x}+2(2a+1)e^{2x}+1 \right)} - \\
- \displaystyle\frac{1}{2\sqrt{a}(a+1)^{3/2}} \tanh^{-1}
\left( \displaystyle\frac{2 \sqrt{a(a+1)}}{e^{2x}+2a+1} \right). \\
\end{array} $$

$$\displaystyle\int \displaystyle\frac{dx}{(\cosh^2x + a)^2} =
\frac{1}{a}\left( \displaystyle\int
\displaystyle\frac{dx}{\cosh^2x + a} - \displaystyle\int
\displaystyle\frac{\cosh^2x \ dx}{(\cosh^2x + a)^2} \right).$$

The definite integrals were calculated.

$$\displaystyle\int_{-\infty}^{\infty}\displaystyle\frac{dx}{\cosh^2x
+ a} = \displaystyle\frac{1}{\sqrt{a(a+1)}}\tanh^{-1}\left(
\sqrt{\displaystyle\frac{a}{a+1}} \right).$$

$$\displaystyle\int_{-\infty}^{\infty}\displaystyle\frac{\cosh^2x \
dx}{(\cosh^2x + a)^2} = \displaystyle\frac{1}{a+1} +
\displaystyle\frac{1}{\sqrt{a}(a+1)^{3/2}} \tanh^{-1} \left(
\sqrt{\displaystyle\frac{a}{a+1}} \right).$$

$$\displaystyle\int_{-\infty}^{\infty}\displaystyle\frac{dx}{(\cosh^2x
+ a)^2} = \displaystyle\frac{1}{\sqrt{a}(a+1)^{3/2}} \tanh^{-1}
\left( \sqrt{\displaystyle\frac{a}{a+1}} \right) -
\displaystyle\frac{1}{a(a+1)}.$$

\section*{Appendix B}

\hspace*{\parindent}There are the expressions for the function
$k(\lambda)$:

\begin{description}
    \item[a)]if $\lambda > 0$, then $k(\lambda) = \sqrt{2 \lambda +1} +
\displaystyle\frac{1}{\sqrt{2 \lambda}} \tanh^{-1}\left(
\sqrt{\displaystyle\frac{2 \lambda}{2 \lambda + 1}} \right)$;
    \item[b)]if $\lambda = 0$, then $k(\lambda) = 2$;
    \item[c)]if $-0.5 <\lambda < 0$, then $k(\lambda) = \sqrt{2 \lambda +1} +
\displaystyle\frac{1}{\sqrt{- 2 \lambda}} \arctan\left(
\sqrt{\displaystyle\frac{- 2 \lambda}{2 \lambda + 1}} \right)$.
\end{description}

% ----------------------------------------------------------------

\end{document}